\documentclass[english]{achemso}
\usepackage[T1]{fontenc}
\usepackage[latin9]{inputenc}
\usepackage{amstext}
\usepackage{amssymb}
\usepackage{graphicx}
\usepackage{esint}

\makeatletter

\title{Photo-thermal 2D spectroscopy:\\ a different type of action}

\author{Pavel Mal\'y$^{1}$}

\email{pavel.maly@matfyz.cuni.cz}

\author{Xiaoji G. Xu$^{2}$}

\author{Tom\'a\v{s} Man\v{c}al$^{1}$}

\affiliation{$^{1}$Faculty of Mathematics and Physics, Charles University, Prague,
Czech Republic\\$^{2}$Department of Chemistry, Stony Brook University,
Stony Brook, New York, United States }

\usepackage{natbib}
\setcitestyle{numbers,square}

\makeatother

\usepackage{babel}
\begin{document}
\begin{abstract}
Advances in multidimensional spectroscopy have seen the rise of action
detection, where coherent response is encoded into incoherent signals.
These are typically proportional to the excited-state population,
an example in 2D electronic spectroscopy (2DES) is fluorescence-detected
2DES (F-2DES), and in 2D infrared (2DIR) tag-loss 2DIR (TL-2DIR).
Very recently, a new type of photo-thermal action signal has been
introduced in 2DIR, detecting the generated heat by atomic force microscopy-based
2DIR (AFM-2DIR). We present a unified theoretical framework for population-
and heat-based 2D spectroscopy, highlighting their complementary features.
In the infrared, TL-2DIR reflects the system\textquoteright s linear
response, whereas AFM-2DIR produces spectra resembling conventional
2DIR, with sensitivity to anharmonicity and mode coupling. Our model
reproduces key experimental AFM-2DIR features, confirming measurement
in a highly nonlinear regime. Extending photothermal detection to
electronic spectroscopy, we compare F-2DES with proposed photo-thermal
2DES (PT-2DES). PT-2DES closely resembles conventional 2DES, while
enjoying the advantages of action detection.

\end{abstract}

\paragraph*{Introduction: Taking action}

Two-dimensional spectroscopy is an established tool to fully resolve
the coherent nonlinear response of the vibrations in the infrared
(2DIR) and the electronic states in the visible (2DES).\cite{jonas_two-dimensional_2003,cho_coherent_2008}
2DIR allowed precise determination of vibrational mode anharmonicity
and inter-mode coupling \cite{hamm_concepts_2011}, with applications
to protein structure \cite{strasfeld_strategies_2009} and way beyond.
\cite{kim_applications_2009,hunt_2d-irspectroscopy_2009,baiz_celebrating_2024}
In 2DES, the coupling between electronic transitions can be determined,
quantifying exciton delocalization and energy transport.\cite{cho_coherent_2008,brixner_two-dimensional_2005}
Examples include complete determination of energy flow in photosynthetic
complexes\cite{rouxel_charting_2026} and beyond. 

The recent decade has seen a rapid rise of the so-called action-detected
2D spectroscopy, which encodes the coherent 2D response into an incoherent
observable. In 2DES, the observable can be fluorescence, \cite{agathangelou_phase-modulated_2021,goetz_coherent_2018,tekavec_fluorescence-detected_2007}
photo-current \cite{bakulin_ultrafast_2016,uhl_coherent_2021,karki_coherent_2014,lopez-ortiz_photoelectrochemical_2024}
or photo-ions.\cite{roeding_coherent_2018,bruder_coherent_2018,solowan_direct_2022}
In 2DIR this approach has emerged only recently, detecting loss of
an attached tag in gas phase\cite{ma_two-dimensional_2023,dutta_nonlinear_2025},
or heat-induced expansion in AFM microscopy\cite{xie_fourier-transform_2022,xie_atomic-force-microscopy-based_2024}.
The action detection has expanded the applicability range and sensitivity
of 2D spectroscopy, isolating the resonant response only, providing
detection free from the excitation pulse background, and facilitating
its application in microscopy with high spatial resolution and sensitivity.\cite{tiwari_spatially-resolved_2018,jana_fluorescence-detected_2024} 

While the action-detection probes in principle the same four-wave
mixing response as the coherently detected signal, its different mode
of detection alters the information content of the spectra. The most
differences arise from the interaction with an additional laser pulse,
adding an excitation pathway contributing to the nonlinear signal.
This excitation pathway ends in a double excited state, whose evolution
during signal emission determines the relative weight of the excited-state
absorption type pathway. This has important consequences for the technique
sensitivity to mode anharmonicity, transition coupling and excitation
dynamics. In electronic spectra of extended systems, the interaction
of two otherwise independent excitations during signal emission leads
to a profound challenge for action spectroscopy known as incoherent
mixing, producing static correlations that obscure the desired nonlinear
response\cite{gregoire_incoherent_2017,bolzonello_nonlinear_2023}.

In this context, photothermal 2D spectroscopy is a different kind
of action since energy deposited in form of excitation does tend to
eventually convert to heat, be it by excited-state relaxation or due
to recombinative interaction of excitations. As a result, photothermal
2D spectroscopy retains the sensitivity of standard coherently detected
2D spectroscopy, while keeping the advantages of action detection.
In this work, we explicitly compare the heat-based photo-thermal 2D
spectra, such as AFM-2DIR, with population-based 2D spectra, such
as TL-2DIR. 

\begin{figure}[h]
\centering
\includegraphics[scale=0.3]{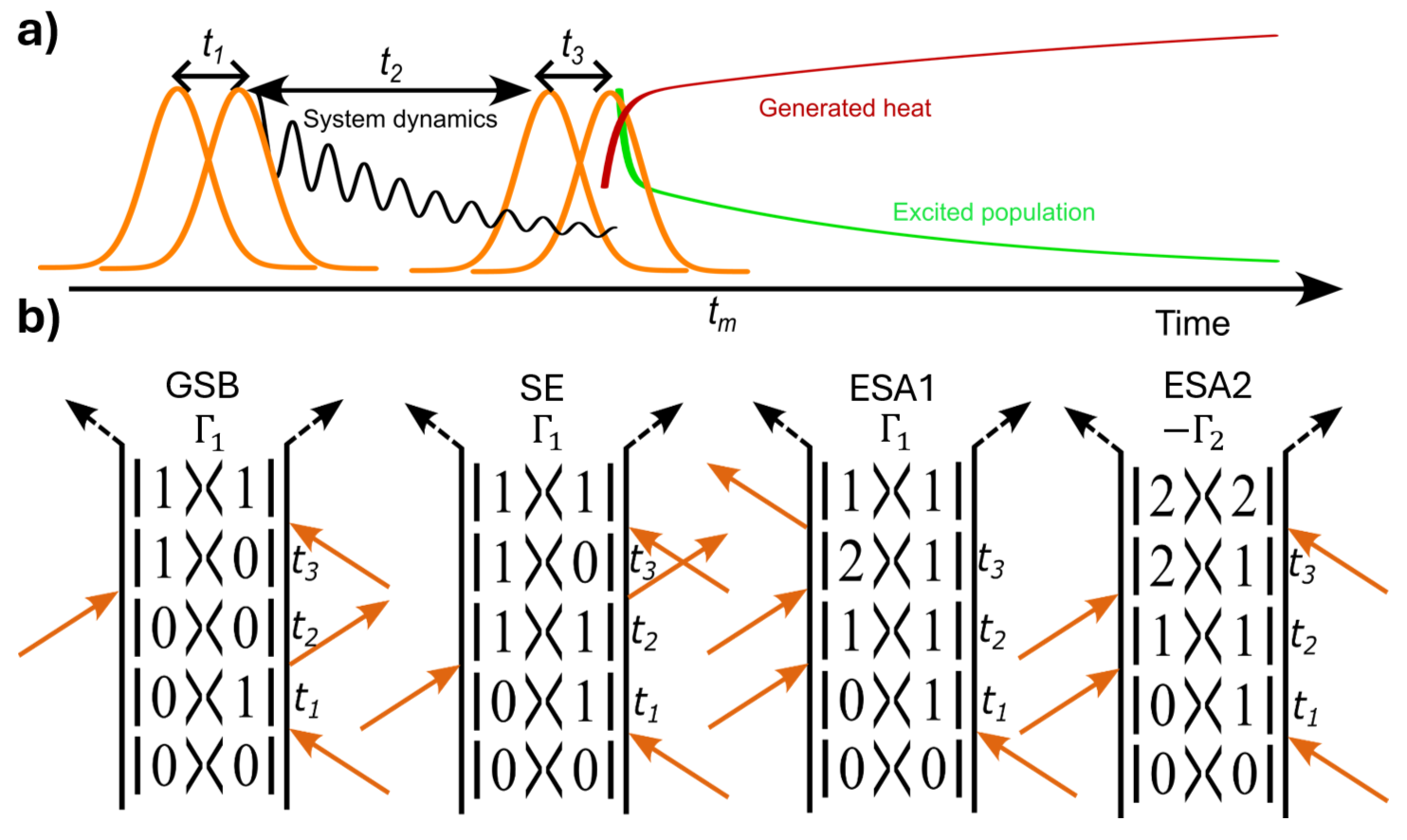}

\caption{a) A typical pulse sequence for action-detected 2D spectroscopy. The
first pulse pair (separated by an interferometric time delay $t_{1}$)
excites the system, which evolves in time $t_{2}$. A second pulse
pair (separated by an interferometric delay $t_{3}$) then probes
the state of the system, creating nonlinear excited-state population.
This population decays, being dissipated as heat or emitted as fluorescence.
b) Four rephasing excitation pathways the system follows being perturbed
by the pulses in a). From the left, there are ground-state-bleach
(GSB) type pathway, stimulated emission (SE) type pathway, and two
excited-state absorption (ESA) pathways. Pathways are depicted by
double-sided Feynman diagrams following the system density matrix
elements evolving in the respective delay times. Solid arrows indicate
interaction with the pulses, dashed arrows emission of the incoherent
signal, time flows upwards.\cite{hamm_concepts_2011} Without considering
the detailed states involved, the overall pathway contributions can
be quantified by factors of $\Gamma_{1}$ and $\Gamma_{2}$, used
in the discussion in the text.\label{fig:A-typical-pulse-sequence}}

\end{figure}

\paragraph*{2D spectroscopy detected by population and by heat}

The system interacting with the excitation pulses (electric field
$E(t)$) can be described by its density matrix $\rho(t)$, evolving
according to the equation of motion\cite{leonas_valkunas_darius_abramavicius_tomas_mancal_molecular_2013}

\begin{equation}
\frac{d\rho(t)}{dt}=-\frac{i}{\hbar}\left[H_{0},\rho(t)\right]-{\cal R}\rho(t)+\frac{i}{\hbar}\left[\mu,\rho\right]E(t).\label{eq:EOM}
\end{equation}
Here $H_{0}$ is the system Hamiltonian of the relevant probed transitions,
$\mu\cdot E(t)$ is interaction with light in semiclassical approximation
($\mu$ is the dipole moment operator), and ${\cal R}$ is the relaxation
super-operator (in principle time-dependent), reflecting dephasing
and energy dissipation due to the interaction with the environment.
In coherently detected 2D spectroscopy, the coherent polarization
response of the system to the interaction with three excitation pulses
is detected. In the action-detected spectroscopy, the incoherent signal
is produced during and after the interaction with the pulses. The
different types of action signals result in different sensitivity
to different properties of the system. The first type of action signals
are proportional to the excited-state population. For typical signals
a separation of timescales is possible in which the signal emission
is slower than the interpulse time interval $\left(t_{0},t_{m}\right)$.
The population-detected 2D signal $S_{\text{pop-2D}}$ can then be
calculated as 
\begin{equation}
S_{\text{pop-2D}}\propto\left(1-\rho_{gg}(t_{m})\right),\label{eq:pop_signal}
\end{equation}

\noindent where $t_{m}$ is a chosen time well after the interaction
with the last pulse of the excitation sequence (see Fig. \ref{fig:A-typical-pulse-sequence}a)
and $\rho_{gg}$ is the population remaining in the ground state.
For 2DIR, such signal can be the TL-2DIR, where the tag is dissociated
whenever the system ends up in a state with energy larger than the
dissociation threshold.\cite{ma_two-dimensional_2023,dutta_nonlinear_2025}
Oftentimes, this state lies under the first excited state, so that
the signal is produced whenever the system ends up excited after the
pulses, regardless of the excitation energy \cite{ma_two-dimensional_2023}.
For 2DES, a typical population-based signal is the fluorescence (F-2DES),
where spectrally unresolved emission reports on the excited state
occupation, and any excitation to a higher excited state does not
lead to more signal (Kasha rule). 

A second kind of action signal we will discuss is proportional to
the dissipated heat induced by the excitation. The signal can thus
be calculated as proportional to the system energy after the excitation
pulses, 

\begin{equation}
E_{\text{sys}}=\text{Tr}\left\{ H\rho(t_{m})\right\} ,
\end{equation}

\noindent plus the energy dissipated during the action of the pulses:

\begin{equation}
E_{\text{diss}}=\int_{t_{0}}^{t_{m}}\text{Tr}\left\{ H\left(\frac{\partial\rho}{\partial t}\right)_{\text{diss}}\right\} dt=\int_{t_{0}}^{t_{m}}\text{Tr}\left\{ H{\cal R}_{\text{diss}}\rho(t)\right\} dt,
\end{equation}

\noindent where ${\cal {\cal R}_{\text{diss}}}$ is the dissipative
part of the relaxation superoperator. The total dissipated energy
is thus

\begin{equation}
S_{\text{PT-2D}}\propto\int_{t_{0}}^{t_{m}}\text{Tr}\left\{ H{\cal R}_{\text{diss}}\rho(t)\right\} dt+\text{Tr}\left\{ H\rho(t_{m})\right\} .\label{eq: end energy signal}
\end{equation}

In the infrared, the photo-thermal detection has been recently realized
by AFM microscopy (AFM-2DIR) measuring the heat-induced sample expansion.\cite{xie_fourier-transform_2022,xie_atomic-force-microscopy-based_2024}
In the visible, PT-2DES has not been realized yet, but we are presently
taking steps in this direction, both using AFM-based and optical detection
approaches. 

\paragraph*{2D spectra calculation}

To directly compare the photo-thermal 2D spectra to the standard coherently
detected 2D, we first consider a four-pulse experiment. To calculate
the 2D spectra, we explicitly integrate the equations of motion for
the reduced density matrix of the system interacting with the electric
field of the pulses, Eq. (\ref{eq:EOM}). The calculations are carried
out by a custom script\footnote{The script will be made openly available upon publication in a Zenodo
repository} using our open-source python-based package quantarhei, available
from GitHub.\cite{mancal_quantarhei_2026} To isolate the nonlinear
contributions corresponding to the the absorptive 2D spectra, we use
phase cycling of the pulses.\cite{tan_theory_2008} Specifically,
we employ the standard 1x3x3x3 27-fold phase cycling, with phase signatures
of {[}-1,+1,-1,+1{]} for the rephasing and {[}+1,-1,+1,-1{]} for the
non-rephasing contribution, adding them together to obtain absorptive
2D spectra. The pulses of length $\tau_{\text{pulse}}$ are centered
resonant with the central frequency of the system $\omega_{0}$. The
scanned delays between the pulses are the coherence times $t_{1}$
and $t_{3}$ and the waiting time $t_{2}$. To speed up the 'numerical
acquisition', the delays are scanned in the so-called partial rotating
frame, i.e., a pulse delayed by $\tau$ gets an additional phase of
$-\left(1-\gamma_{\text{rf}}\right)\omega_{0}\tau$, where $\gamma_{\text{rf}}=1$
means the laboratory frame (no rotation) and $\gamma_{\text{rf}}=0$
is the fully rotating frame.\cite{maly_coherently_2020} The rotating
frame shifts the signals oscillating in $t_{1}$ and $t_{3}$ at a
frequency around $n\omega_{0}$ to the frequency $n\gamma\omega_{0}$,
enabling us to use larger coherence time steps for the scans. We used
$\gamma_{\text{rf}}=0.05$ in the IR and $\gamma_{\text{rf}}=0.1$
in the visible. The step size is given by the maximum frequency $\omega_{\text{max}}$
to be sampled via the Nyquist criterion, we take $\omega_{\text{max}}=\omega_{0}\gamma_{\text{rf}}+1.5\frac{2\pi}{\tau_{\text{pulse}}}$
to cover the spectral bandwidth.  The system dynamics is calculated
in the rotating wave approximation ($H_{\text{int}}=-\mu^{+}E^{-}-\mu^{-}E^{+}$),
using Lindblad superoperator formalism for relaxation and dephasing.\cite{leonas_valkunas_darius_abramavicius_tomas_mancal_molecular_2013}
The propagation is done in the rotating frame (around center frequency
$\omega_{0}$), allowing us to use larger time steps $\frac{\tau_{\text{pulse}}}{20}$.
The signals are calculated from the time-dependent density matrix
using Eq. (\ref{eq:pop_signal}) for the population-based signal and
Eq. (\ref{eq: end energy signal}) for the photo-thermal signal. In
addition, the linear excitation spectra are calculated using a pulse
pair with interferometric delay $t_{1}$. Notice, that the calculation
formulated like this can be equally well carried out in the infrared
as well as in the visible. To interpret the signals, we will discuss
the 2D spectra in terms of excitation pathways, shown in Fig. (\ref{fig:A-typical-pulse-sequence})b.
Note, however, that the calculations are done non-perturbatively,
with realistic pulses and all possible light-matter interactions. 

\begin{figure}[h]
\centering
\includegraphics[scale=0.3]{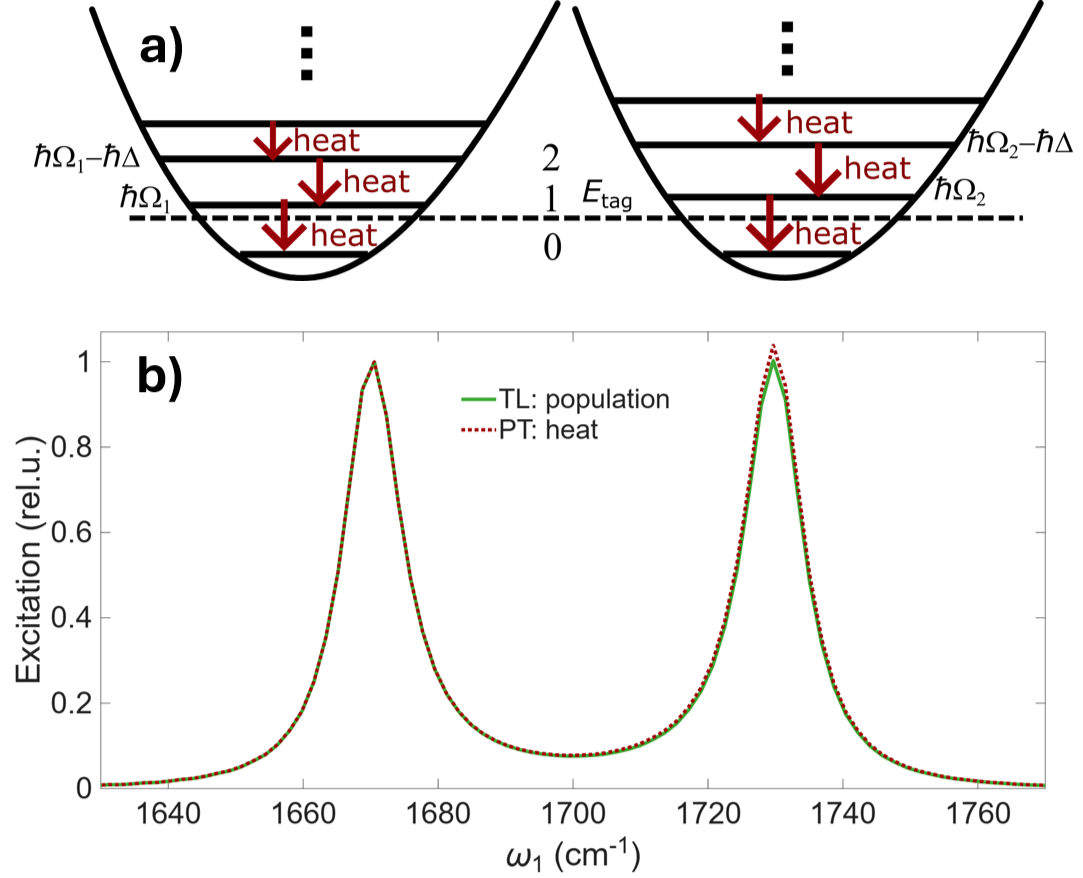}

\caption{a) Two anharmonic vibrational modes used as a model system in the
calculation. The energy $E_{\text{tag}}$ needed for tag release is
denoted by the dashed horizontal line, the dissipated heat due to
vibrational relaxation is marked by red arrows. Numbers in the center
denote the general multiple-excitation manifolds, as used in the excitation
pathways in Fig. \ref{fig:A-typical-pulse-sequence}b. b) The linear
excitation spectrum calculated by a double-pulse scan, with the population-based
(green solid) and heat-based (red dashed) detection.\label{fig:Vibstates}}

\end{figure}

\paragraph*{TL-2DIR vs AFM-2DIR}

2DIR probes characteristic vibrational modes, their anharmonicity
and coupling. We therefore consider two coupled (an)harmonic modes
as a model system, sketched in Fig. \ref{fig:Vibstates}a. For convenience
and to avoid pulse-spectrum distortions, we set the mode frequencies
symmetrically around the laser frequency $\omega_{0}=1700\text{ cm}^{-1},$at
$[\Omega_{1},\Omega_{2}]=[1670,1730]$ cm$^{-1}$. We will take the
modes to be either anharmonic with anharmonicity $\Delta=20\text{ cm}^{-1}$
or harmonic with $\Delta=0\text{ cm}^{-1}$. The modes can be uncoupled,
or coupled with interaction strength $V=20\text{ cm}^{-1}$, and the
coupling takes the bi-linear form in the mode coordinates, $V=Vq_{1}q_{2}=\frac{V}{2}\left(b_{1}^{\dagger}+b_{1}\right)\left(b_{2}^{\dagger}+b_{2}\right).$
As for the standard harmonic oscillator, we will consider the transition
oscillator strength to increase with the quantum number $\mu_{n+1,n}=\sqrt{n+1}\mu_{0},$
and the same pure dephasing for all transitions, $\gamma_{\text{pd}}^{n,n-1}=\gamma_{\text{pd}}^{10}=\gamma_{\text{pd}}=1\text{ ps}^{-1}$.
Since we discuss zero-waiting-time spectra only, we will for simplicity
neglect vibrational relaxation during the propagation, which would
only broaden the lineshapes by additional lifetime-induced dephasing.\cite{hamm_concepts_2011}
 For excitation, we use a train of four Gaussian pulses, $\tau_{\text{pulse}}=200\text{ fs}$.
First, we calculate the excitation spectra using a double-pulse scan,
finding the expected two peaks of the two vibrational modes, see Fig.
\ref{fig:Vibstates}b. The spectra are scaled on the lower-energy
peak, the relative amplitude of the higher-energy peak is slightly
larger for the heat-based detection since excitation of the higher-frequency
mode produces slightly (about 3.5\%) more heat compared to the lower
peak. This is a general effect in the photothermal detection, where
higher-energy peaks are emphasized, but unless the spectral bandwidth
is very large, this effect is small. 

\begin{figure}[t]
\centering
\includegraphics[scale=0.4]{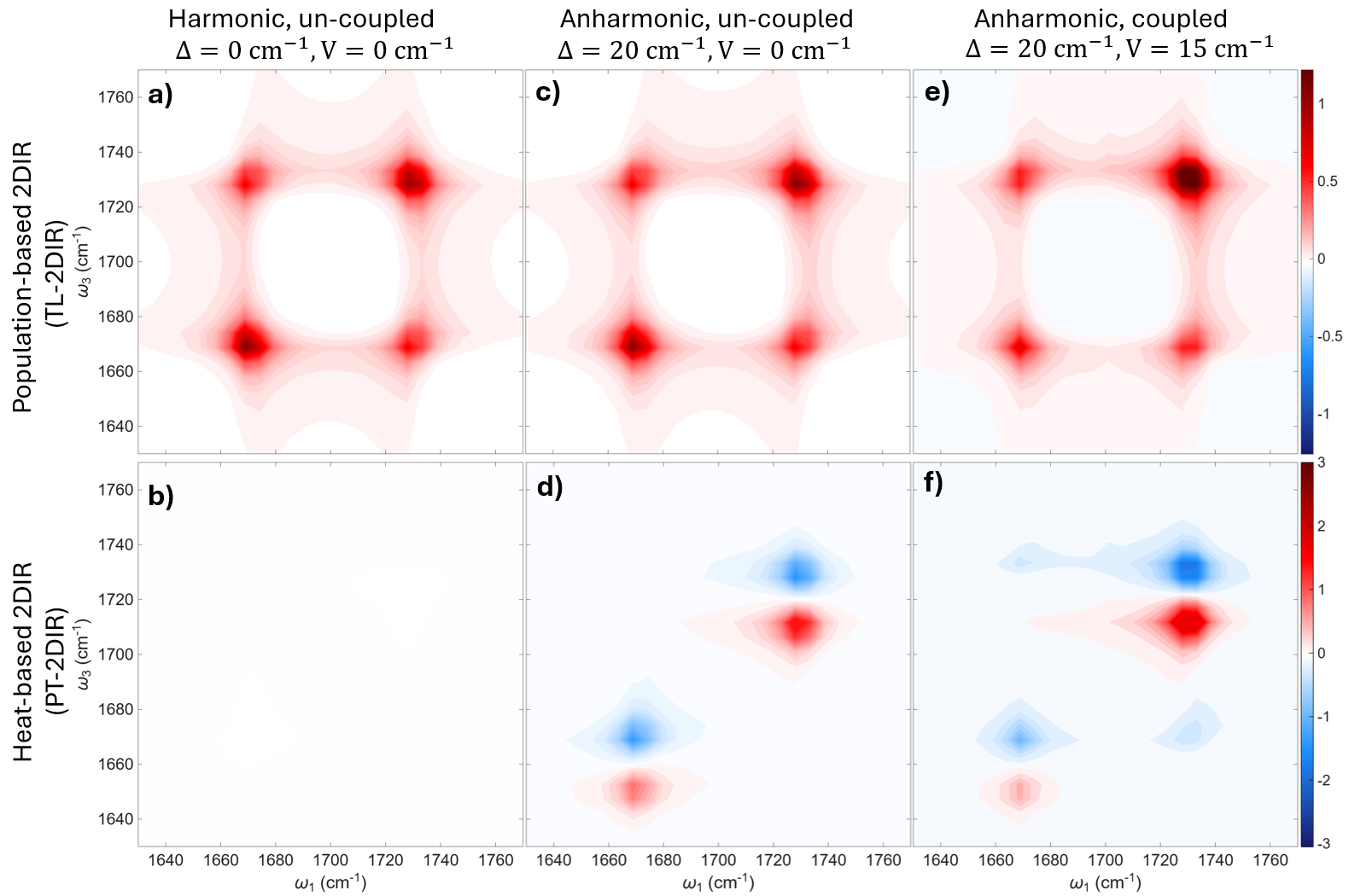}

\caption{Absorptive TL-2DIR (top, a-c) and AFM-2DIR (bottom, d-f). Shown are
the spectra of two modes with varying anharmonicity $\Delta$ and
coupling strength $V$. Note the identical color scale for all TL-2DIR
spectra (a,c,e) and for all AFM-2DIR spectra (b,d,f). \label{fig:RephNonreph_spectra-dimer}}
\end{figure}

The calculated 2D spectra are shown in Fig. \ref{fig:RephNonreph_spectra-dimer},
for varying mode coupling and anharmonicity. First, consider uncoupled
harmonic modes with $V=0\text{ cm}^{-1}$, $\Delta=0\text{ cm}^{-1}$,
Fig. \ref{fig:RephNonreph_spectra-dimer}a,b. The TL-2DIR spectrum,
Fig. \ref{fig:RephNonreph_spectra-dimer}a, features two diagonal
peaks at the corresponding frequencies, as well as cross peaks indicating
spectral correlations resulting from the fact that simultaneous excitation
of both modes still leads to a release of a single tag. In contrast,
the AFM-2DIR spectrum, Fig. \ref{fig:RephNonreph_spectra-dimer}b,
shows no signal for harmonic modes, same as in 2DIR. To interpret
the 2DIR spectra, we can use the language of excitation pathways shown
in Fig. \ref{fig:A-typical-pulse-sequence}b, whether the pathway
contributions of $\Gamma_{1}$ and $\Gamma_{2}$ are defined. The
standard, coherently detected 2DIR spectrum can be expressed as $\text{2DIR}=\text{GSB}+\text{SE}-\text{ESA}$.
For TL-2DIR, the pathways contribute by one released tag, regardless
of the end state, so that $\Gamma_{1}=\Gamma_{2}$ and $\text{TL-2DIR}=\text{GSB}+\text{SE}$.
The ESA is absent since the induced absorption does not lead to an
increased probability of tag release. Note, that this would change
if one would engineer a tag with the release energy above the first
excited state, in which case one would have $\Gamma_{1}=0$ and $\Gamma_{2}=1$,
measuring exclusively the ESA. Unless this is the case, the TL-2DIR
spectrum consists of the GSB and SE pathways only. In PT-2DIR, we
have $\Gamma_{1}=\hbar\omega_{10}$ and $\Gamma_{2}=\hbar\omega_{10}+\hbar\omega_{21}$,
so that the total signal is $\text{PT-2DIR}=\hbar\omega_{10}\left\{ \text{GSB}+\text{SE}-\text{ESA}\frac{\omega_{21}}{\omega_{10}}\right\} $.
For weak anharmonicity, $\frac{\omega_{21}}{\omega_{10}}=1-\frac{\Delta}{\omega_{10}}\approx1$
and the PT-2DIR spectrum approaches the standard 2DIR. 

The interference of the excitation pathways makes standard 2DIR sensitive
to the mode anharmonicity: for harmonic modes, the excitation pathways
destructively interfere and the spectrum vanishes. Mathematically,
for a single mode the spectrum can be expressed by $\text{AFM-2DIR}=2|\mu_{10}|^{4}{\cal L}_{10}(\omega_{1})\left\{ {\cal L}_{10}(\omega_{3})-{\cal L}_{21}(\omega_{3})\right\} $,
with ${\cal L}_{n+1,n}(\omega)$ representing the lineshape of the
respective transition. For a harmonic mode, ${\cal L}_{10}(\omega_{3})={\cal L}_{21}(\omega_{3})$,
the ESA exactly cancels the GSB+SE and the peaks vanish. Anharmonic
modes then feature the characteristic dispersive lineshape, with peak
separation along $\omega_{3}$ reflecting the mode anharmonicity $\Delta$.\cite{hamm_concepts_2011}
For TL-2DIR the ESA does not contribute, so one gets $\text{TL-2DIR}=2|\mu_{10}|^{4}{\cal L}_{10}(\omega_{1}){\cal L}_{10}(\omega_{3})$,
which is a product of linear lineshapes along the two axes. This is
why the Reppert group recently called TL-2DIR a ``nonlinear response
from linear oscillators''.\cite{dutta_nonlinear_2025} Accordingly,
introducing anharmonicity $\Delta=20\text{ cm}^{-1}$ does not change
TL-2DIR spectrum, Fig. \ref{fig:RephNonreph_spectra-dimer}c. In contrast,
in PT-2DIR (Fig. \ref{fig:RephNonreph_spectra-dimer}d) the two modes
appear with the characteristic lineshapes, contributing to the heat
signal independently same as in standard 2DIR.

Finally, let us consider coupled modes, $V=15\text{ cm}^{-1}$. The
immediate effect of the finite coupling is the redistribution of the
oscillator strength of the eigenstates, visible both in TL-2DIR (Fig.
\ref{fig:RephNonreph_spectra-dimer}e) and AFM-2DIR (Fig. \ref{fig:RephNonreph_spectra-dimer}f)
on the relative amplitude of the diagonal peaks. In TL-2DIR, the coupling
barely affects the cross peaks. In contrast, in AFM-2DIR the cross
peaks appear only with some finite coupling. Overall, same as in 2DIR,
the PT-2DIR cross peaks measure the mode coupling, and the presence
of the peaks indicates mode anharmonicity. 

\begin{figure}[t]
\centering
\includegraphics[scale=0.4]{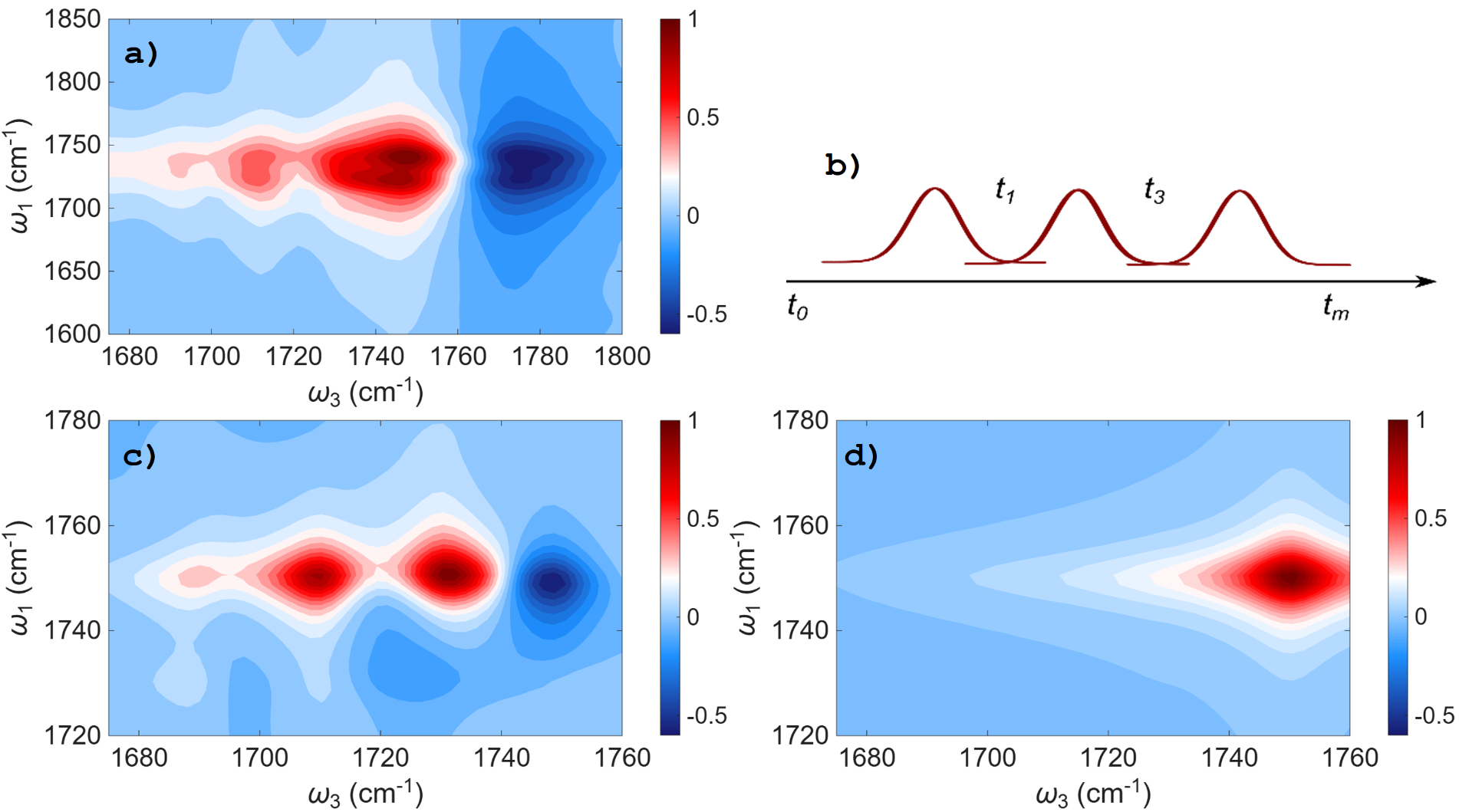}\caption{AFM-2DIR of a carbonyl vibrational mode, experiment and theory. a)
Experimental AFM-2DIR spectrum. b) Measurement three-pulse sequence.
c) Calculated AFM-2DIR spectrum for high-intensity excitation. d)
Calculated AFM-2DIR spectrum for low-intensity excitation. \label{fig:Carbonyl_mode_exp_theo}}
\end{figure}

\paragraph*{Comparison to experimental AFM-2DIR: Three-pulse experiments}

The pioneering AFM-2DIR experiments, see spectrum in Fig. \ref{fig:Carbonyl_mode_exp_theo}a,
were carried out using three-pulse sequences, see Fig. \ref{fig:Carbonyl_mode_exp_theo}b,
without any phase cycling\cite{xie_atomic-force-microscopy-based_2024}.
Since this is experimentally much simpler to realize and facilitates
significantly faster data acquisition, we investigate the PT-2DIR
signals acquired with the corresponding pulse sequences. The experimental
AFM-2DIR spectrum of a carbonyl vibrational mode, taken from Ref.
\cite{xie_atomic-force-microscopy-based_2024}, can be found in Fig.
\ref{fig:Carbonyl_mode_exp_theo}a. To honor the experimental plotting
convention, we follow here the same labeling as in the original manuscript,
with the excitation frequency $\omega_{1}$ on the vertical axis and
the detection frequency $\omega_{3}$ on the horizontal (labeled as
$\omega_{2}$ in the original publication, for pulse sequence see
Fig. \ref{fig:Carbonyl_mode_exp_theo}b). The experimental spectrum
shows a single frequency along the excitation axis $\omega_{1}$,
corresponding to the excited mode, and a sequence of peaks along the
detection axis $\omega_{3}$, reflecting highly excited vibrational
ladder of the mode, with the peak spacing given by the anharmonicity.
Note, that the peak position along the excitation axis doesn't precisely
match the 1-0 transition frequency. This can possibly be caused by
the laser spectrum shape or other experimental factors such as interferogram
scan calibration, which we disregard here. Our goal is to show qualitative
agreement with the theory, without detailed fitting of the system
parameters. For the calculation, we thus simply consider a single
anharmonic mode ($\Omega=1750\text{ cm}^{-1},$ $\Delta=20\text{ cm}^{-1},\gamma_{\text{pd}}=1\text{ps}^{-1}$).
Following the data processing in the experimental work\cite{xie_atomic-force-microscopy-based_2024},
we first take Fourier transform of the real data $S(t_{1},t_{3})$
in the second coherence time (here $t_{3}$, there $t_{2}$) for each
$t_{1}$, getting $S(t_{1}$,$\omega_{3})$. Then, we take the absolute
part and carry out the additional Fourier transform over $t_{1}$,
$S(\omega_{1},\omega_{3})={\cal F}_{1}|S(t_{1},\omega_{3})|$. Due
to the field enhancement by the AFM tip, we consider a relatively
strong excitation, leading to 30\% excited-state population after
the action of the three pulses (at maximum pulse time delay, and reachign
80\% at full pulse overlap, way beyond any perturbative regime). The
resulting spectrum is shown in Fig. \ref{fig:Carbonyl_mode_exp_theo}c.
Clearly, the qualitative agreement with the experiment is good, including
peak amplitudes and shapes. The probe finds the vibrational mode highly
excited, manifesting as a bleach of the 0-1 transition, and induced
absorption of the higher transitions shifted by multiples of the anharmonicity
$\Delta$. As already anticipated from the highly-excited-state progression,
the high intensity with multiple subsequent interactions with the
excitation pulse pair is necessary, in line with a very recent perturbative
calculation by the Kananenka group.\cite{rueda_espinosa_theory_2026}
In the low-intensity regime, the spectrum would be given by the linear
response to the interfering first and last pulses, as seen in Fig.
\ref{fig:Carbonyl_mode_exp_theo}d. To see the origin of this nonlinear
response, consider that the 2D Fourier transform isolates only the
signals oscillating both in $t_{1}$ and $t_{3}$. Among all the linear
signals from the three pulses, only the excitation pathway with one
interaction with the first pulse and one with the last pulse satisfies
this criterion, with the optical coherence between them oscillating
as $\propto e^{\pm i\omega_{10}\left(t_{1}+t_{3}\right)}$. In the
weak-field regime, this linear contribution dominates the data, so
that only a single peak at $\omega_{10}$ along both axes is seen
in Fig. \ref{fig:Carbonyl_mode_exp_theo}d.

The high excitation pulse intensity makes it possible to observe the
nonlinear response without isolation of a particular nonlinear signal
contribution. For more complex systems than a single mode, however,
the interpretation of the results would quickly become cumbersome
and would rely on direct modeling, for which the precise pulse intensity
needs to be known. Taking into account effects such as light spatial
distribution around the tip, the situation can quickly become intractable.
It is therefore desirable to isolate the nonlinear response directly
in the experiment, and optionally restrict oneself to the lowest nonlinear
(i.e. fourth) order. To eliminate the linear response, multiple datasets
must be measured, for which the nonlinear signal changes differently
than the linear one from pulses 1 and 3. The simplest possibility
is subtraction of data with 3 pulses minus the data without the middle
pulse 2. More general possibility are phase cycling schemes\cite{tan_theory_2008},
although one should note that with three pulses only the rephasing
contribution {[}-1,+2,-1{]} phase signature can be isolated, since
the non-rephasing contribution {[}+1,0,-1{]} has the same phase signature
as the linear contribution {[}+1,-1{]}. Finally, it is possible to
use intensity cycling recently developed by some of us.\cite{maly_separating_2023,luttig_simple_2026}
The last option is intriguing since it allows complete signal decomposition
into progressively nonlinear spectra. A combination of phase and intensity
cycling can also be used, for example to remove the linear component
with three pulses and get the non-rephasing spectrum. One must, however,
be cautious, since the non-perturbative regime of the interaction
could easily be reached due to the tip-induced enhancement, in which
case any perturbative intensity-dependent decomposition loses meaning.
The applicability of the perturbative description must be checked
by sufficiently decreasing nonlinear order amplitude. \cite{maly_separating_2023}

\begin{figure}[t]
\centering
\includegraphics[scale=0.3]{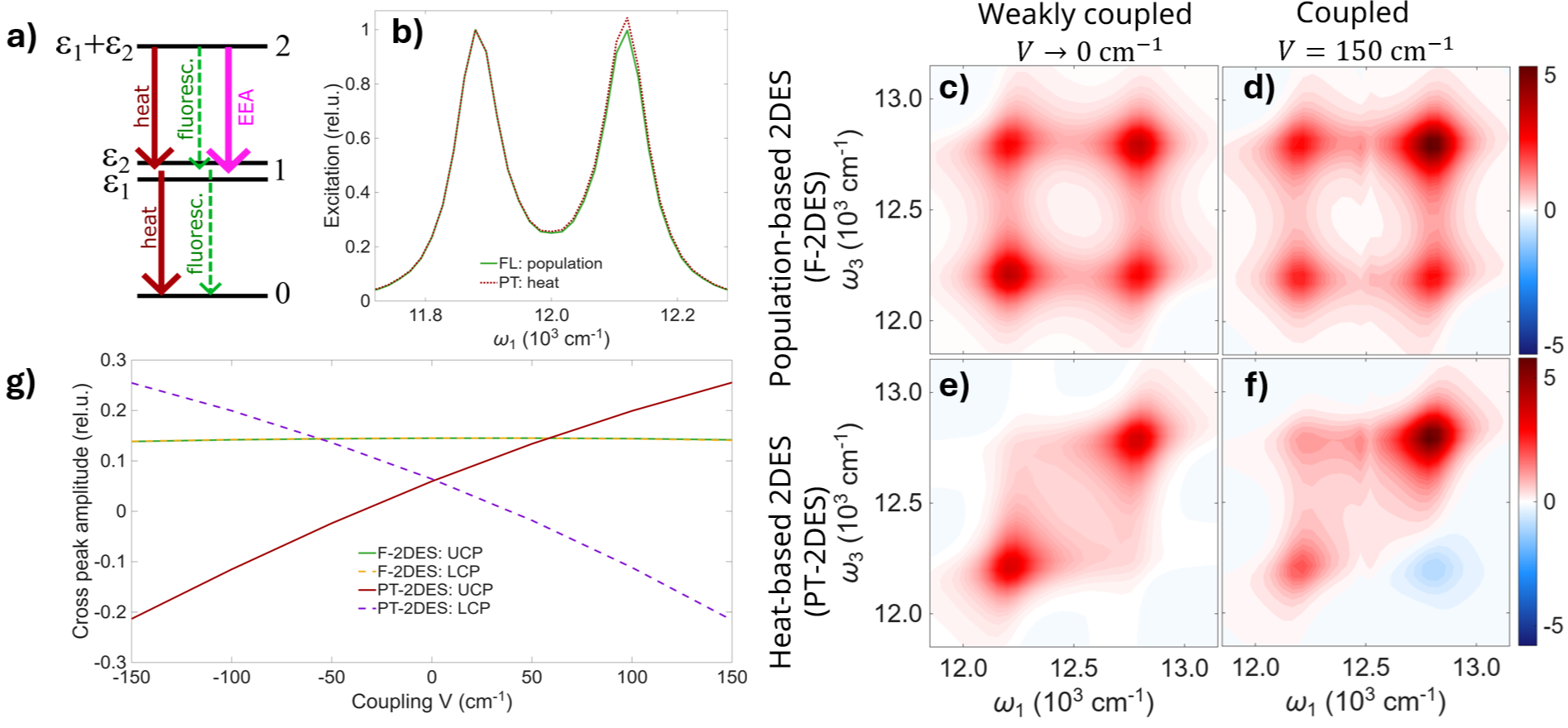}

\caption{a) Energy level scheme of a molecular hetero-dimer used for the calculations,
with relevant processes indicated. b) Linear excitation spectrum with
fluorescence and photo-thermal detection. F-2DES spectra with very
weak coupling (c) and appreciable coupling (d). PT-2DES spectra with
negligible coupling (e) and appreciable coupling (f). g) Cross peak
amplitude dependence on the electronic coupling for F-2DES and PT-2DES,
LCP: lower cross peak, UCP: upper cross peak. \label{fig:2DES}}
\end{figure}

\paragraph*{PT-2DES vs F-2DES}

Having established the properties of PT-2DIR, we now proceed to the
visible spectral range, where electronic transitions are probed. As
a system, we consider a coupled molecular heterodimer, with transition
energies $\left[\epsilon_{1},\epsilon_{2}\right]=\left[12200,12800\right]\text{ cm}^{-1}$,
see Fig. \ref{fig:2DES}a. The same explicit calculation as we used
in the infrared can be used to calculate the 2D spectra in the visible.
We take the pure dephasing to be $\gamma_{\text{pd}}=0.02\text{ ps}^{-1}$
and excite the system by 15 fs pulses centered at 12500 cm$^{-1}$.
The excitation spectra obtained by a two-pulse scan (delay $t_{1}$),
Fig. \ref{fig:2DES}b, show two peaks corresponding to the two molecules.
Normalizing on the lower-energy peak, we again notice the slightly
higher amplitude of the upper peak in photothermal detection since
the excitation of the higher-energy transition produces about 5\%
more heat. Same as in 2DIR, the higher-energy peaks are thus slightly
emphasized. 

We directly proceed with the 2D spectra. A typical population-based
2DES variant is fluorescence-detected F-2DES, whose interpretation
has been discussed by us and others previously \cite{maly_signatures_2018,kunsel_simulating_2019}.
Because of the Kasha rule, the fluorescence is emitted from the lowest
excited state (or excited-state quasi-stationary thermal equilibrium).
For the excitation pathways, defined in Fig. \ref{fig:A-typical-pulse-sequence}b,
we therefore have weights $\Gamma_{1}=\phi_{\text{fl}}$, $\Gamma_{2}=\phi_{\text{fl}}\left(1+\phi_{\text{rel}}^{(2)}\right)$,
where $\phi_{\text{fl}}$ is the fluorescence quantum yield and $\phi_{\text{rel}}^{(2)}$
is the fluorescence yield of state 2 relative to 1, taking into account
additional recombination channels such as exciton--exciton annihilation
(EEA) or fast internal conversion. The F-2DES spectrum is thus $\text{F-2DES}=\phi_{\text{fl}}\left\{ \text{GSB}+\text{SE}-\phi_{\text{rel}}^{(2)}\text{ESA}\right\} .$
If there were no EEA and the double excited state 2 produced two times
more photons than state 1, $\phi_{\text{rel}}^{(2)}\rightarrow1$
and F-2DES would be equivalent to 2DES. In connected systems such
as our dimer, however, EEA is very efficient, so that only one photon
is emitted from a double-excited state and $\phi_{\text{rel}}^{(2)}\rightarrow0$.
As a result, the ESA pathways cancel same as in TL-2DIR, $\text{F-2DES}=\phi_{\text{fl}}\left(\text{GSB}+\text{SE}\right)$.
The EEA has an opposite role in PT-2DES, where the resulting heat
produces signal. In contrast to 2DIR, there is, however, a loss channel
for the PT-2DES: the fluorescence. The excitation pathway weights
are thus $\Gamma_{1}=\left(1-\phi_{\text{fl}}\right)$ and $\Gamma_{2}=2-\phi_{\text{fl}}\left(1+\phi_{\text{rel}}^{(2)}\right)$.
The PT-2DES signal is thus $\text{PT-2DES}=\left(1-\phi_{\text{fl}}\right)\left\{ \text{GSB}+\text{SE}-\frac{1-\phi_{\text{fl}}\phi^{(2)}}{1-\phi_{\text{fl}}}\text{ESA}\right\} $.
For weakly fluorescent samples $\phi_{\text{fl}}\rightarrow0$ and
$\text{PT-2DES}=\text{GSB}+\text{SE}-\text{ESA}$ thus corresponds
to the 2DES, same as PT-2DIR resembles 2DIR, regardless of the amount
of exciton--exciton annihilation. For strongly fluorescent samples,
the PT-2DES signal depends on the extent of EEA. For no EEA, the PT-2DES
signal equals the 2DES, in this situation we have, interestingly,
PT-2DES=F-2DES=2DES. For efficient EEA, though, more heat will come
from the double excited states breaking the balance of the pathways,
yielding $\text{PT-2DES}=\left(1-\phi_{\text{fl}}\right)\left\{ \text{GSB}+\text{SE}-\frac{1}{1-\phi_{\text{fl}}}\text{ESA}\right\} .$
In this situation, the F-2DES and PT-2DES become to a degree complementary,
with F-2DES measuring GSB+SE only, and PT-2DES having all contributions
but with pronounced ESA. For appreciable heat and fluorescence, both
PT-2DES and F-2DES suffer from the incoherent mixing, in F-2DES part
of the GSB contribution (due to EEA, otherwise independent excitons
appear bleached), and in PT-2DES part of ESA (due to EEA, otherwise
independent excitons create more signal). An interesting possibility
in this situation is a simultaneous measurement of F-2DES and PT-2DES,
whose linear combination allows to separate the GSB+SE and ESA parts,
eliminating the incoherent mixing and providing access to excited-state
dynamics. Note, that the double excited state need not be only a two-exciton
state, but possibly a higher excited state of the molecules as well.
In that case, the fast relaxation process is not EEA, but only its
second step, internal conversion. As a result, ESA into higher excited
states is typically invisible in F-2DES but present in PT-2DES.

In the following discussion of 2D spectra, we will consider the case
of efficient EEA and weak fluorescence. The F-2DES spectra for our
dimer are shown in Fig. \ref{fig:2DES}c-d and of PT-2DES in Fig.
\ref{fig:2DES}e-f. We focus here on the sensitivity to electronic
coupling between the molecules. A standard 2DES provides clear signatures
of excitonic delocalization in terms of cross peaks, whose amplitude
directly reflects the magnitude and sign of the electronic coupling.
The sensitivity to coupling results from an interference of GSB and
ESA excitation pathways, as is well known and as we have discussed
in detail previously\cite{maly_signatures_2018}. In F-2DES, Fig.
\ref{fig:2DES}c-d, the absence of the ESA results again in GSB-type
correlation between the two transitions, termed incoherent mixing
in larger systems. The cross peaks are thus indicative of EEA presence
rather than electronic coupling. For PT-2DES, Fig. \ref{fig:2DES}e-f,
the ESA is present and the sensitivity of the cross peaks to the couplings
is therefore the same as in standard 2DES. In Fig. \ref{fig:2DES}g,
we plot the cross peak amplitude dependence on the electronic coupling.
The F-2DES cross peaks are present for all coupling strengths, slightly
decreasing with increasing coupling magnitude, in perfect agreement
with previous results\cite{maly_signatures_2018,maly_coherently_2020}.
In contrast, the PT-2DES cross peaks directly measure the coupling
strength and sign. The slight positive off-set of the cross peak amplitude
is caused by the broad Lorentzian peak shapes, see e.g. Fig. \ref{fig:2DES}e.
For weakly fluorescent samples, PT-2DES features the same sensitivity
to electronic coupling as the standard 2DES. Similar conclusions,
although not discussed in this manuscript, apply for visibility of
excited-state dynamics.

\paragraph*{Conclusion}

We have formulated a theoretical description of the heat-based, photo-thermal
2D nonlinear spectroscopy, within the framework typical for the established
action- and coherently detected 2D spectroscopy. While the calculations
are done by explicit integration of the equations of motion, the spectra
are interpreted in the perturbative language of excitation pathways.
We have directly compared heat-based (photo-thermal) and population-based
2D spectroscopy. In the infrared, the specific examples comprise AFM-2DIR
and TL-2DIR, in the visible PT-2DES and F-2DES, where PT-2DES still
is to be realized experimentally. On exemplary 2D spectra of coupled
vibrational modes, we have demonstrated sensitivity of the PT-2DIR
to mode anharmonicity and coupling, following that of standard 2DIR
and contrasting with TL-2DIR. Verifying the PT-2DIR description, we
have reproduced experimental AFM-2DIR spectra of a carbonyl mode.
Finally, we compared PT-2DES and F-2DES in the visible, where the
situation is somewhat more involved due to the complementary role
of the heat and fluorescence as the loss and signal channels in the
two modes of detection. Nevertheless, we found that under rather common
conditions PT-2DES again retains the sensitivity of 2DES to electronic
coupling and dynamics. At the same time, a potential dual fluorescence
and photo-thermal detection scheme would us to further dissect the
nonlinear response in terms of excitation pathways. Our results thus
describe the photo-thermal detection in PT-2DIR and PT-2DIR as a promising
new type of action, enjoying the advantages of action-based approach
while retaining the sensitivity of the standard coherently detected
signals.

\section*{Acknowledgements}

P.M. acknowledges funding by Charles University (Grant No. PRIMUS/24/SCI/007).
This work was supported by the Czech Science Foundation (GA\v{C}R),
grant no. 26-23570S.

\bibliographystyle{achemso}
\bibliography{PT-2D_bib}

\providecommand{\latin}[1]{#1}
\makeatletter
\providecommand{\doi}
  {\begingroup\let\do\@makeother\dospecials
  \catcode`\{=1 \catcode`\}=2 \doi@aux}
\providecommand{\doi@aux}[1]{\endgroup\texttt{#1}}
\makeatother
\providecommand*\mcitethebibliography{\thebibliography}
\csname @ifundefined\endcsname{endmcitethebibliography}
  {\let\endmcitethebibliography\endthebibliography}{}
\begin{mcitethebibliography}{37}
\providecommand*\natexlab[1]{#1}
\providecommand*\mciteSetBstSublistMode[1]{}
\providecommand*\mciteSetBstMaxWidthForm[2]{}
\providecommand*\mciteBstWouldAddEndPuncttrue
  {\def\EndOfBibitem{\unskip.}}
\providecommand*\mciteBstWouldAddEndPunctfalse
  {\let\EndOfBibitem\relax}
\providecommand*\mciteSetBstMidEndSepPunct[3]{}
\providecommand*\mciteSetBstSublistLabelBeginEnd[3]{}
\providecommand*\EndOfBibitem{}
\mciteSetBstSublistMode{f}
\mciteSetBstMaxWidthForm{subitem}{(\alph{mcitesubitemcount})}
\mciteSetBstSublistLabelBeginEnd
  {\mcitemaxwidthsubitemform\space}
  {\relax}
  {\relax}

\bibitem[Jonas(2003)]{jonas_two-dimensional_2003}
Jonas,~D.~M. Two-dimensional femtosecond spectroscopy. \emph{Annu. Rev. Phys.
  Chem.} \textbf{2003}, \emph{54}, 425--463\relax
\mciteBstWouldAddEndPuncttrue
\mciteSetBstMidEndSepPunct{\mcitedefaultmidpunct}
{\mcitedefaultendpunct}{\mcitedefaultseppunct}\relax
\EndOfBibitem
\bibitem[Cho(2008)]{cho_coherent_2008}
Cho,~M. Coherent two-dimensional optical spectroscopy. \emph{Chemical Reviews}
  \textbf{2008}, \emph{108}, 1331--1418\relax
\mciteBstWouldAddEndPuncttrue
\mciteSetBstMidEndSepPunct{\mcitedefaultmidpunct}
{\mcitedefaultendpunct}{\mcitedefaultseppunct}\relax
\EndOfBibitem
\bibitem[Hamm and Zanni(2011)Hamm, and Zanni]{hamm_concepts_2011}
Hamm,~P.; Zanni,~M. \emph{Concepts and methods of {2D} infrared spectroscopy},
  1st ed.; Cambridge University Press: New York, 2011\relax
\mciteBstWouldAddEndPuncttrue
\mciteSetBstMidEndSepPunct{\mcitedefaultmidpunct}
{\mcitedefaultendpunct}{\mcitedefaultseppunct}\relax
\EndOfBibitem
\bibitem[Strasfeld \latin{et~al.}(2009)Strasfeld, Ling, Gupta, Raleigh, and
  Zanni]{strasfeld_strategies_2009}
Strasfeld,~D.~B.; Ling,~Y.~L.; Gupta,~R.; Raleigh,~D.~P.; Zanni,~M.~T.
  Strategies for {Extracting} {Structural} {Information} from {2D} {IR}
  {Spectroscopy} of {Amyloid}: {Application} to {Islet} {Amyloid}
  {Polypeptide}. \emph{J. Phys. Chem. B} \textbf{2009}, \emph{113},
  15679--15691\relax
\mciteBstWouldAddEndPuncttrue
\mciteSetBstMidEndSepPunct{\mcitedefaultmidpunct}
{\mcitedefaultendpunct}{\mcitedefaultseppunct}\relax
\EndOfBibitem
\bibitem[Kim and Hochstrasser(2009)Kim, and
  Hochstrasser]{kim_applications_2009}
Kim,~Y.~S.; Hochstrasser,~R.~M. Applications of {2D} {IR} {Spectroscopy} to
  {Peptides}, {Proteins}, and {Hydrogen}-{Bond} {Dynamics}. \emph{J. Phys.
  Chem. B} \textbf{2009}, \emph{113}, 8231--8251\relax
\mciteBstWouldAddEndPuncttrue
\mciteSetBstMidEndSepPunct{\mcitedefaultmidpunct}
{\mcitedefaultendpunct}{\mcitedefaultseppunct}\relax
\EndOfBibitem
\bibitem[Hunt(2009)]{hunt_2d-irspectroscopy_2009}
Hunt,~N.~T. {2D}-{IRspectroscopy}: ultrafast insights into biomolecule
  structure and function. \emph{Chem. Soc. Rev.} \textbf{2009}, \emph{38},
  1837--1848\relax
\mciteBstWouldAddEndPuncttrue
\mciteSetBstMidEndSepPunct{\mcitedefaultmidpunct}
{\mcitedefaultendpunct}{\mcitedefaultseppunct}\relax
\EndOfBibitem
\bibitem[Baiz \latin{et~al.}(2024)Baiz, Bredenbeck, Cho, Jansen, Krummel, and
  Roberts]{baiz_celebrating_2024}
Baiz,~C.; Bredenbeck,~J.; Cho,~M.; Jansen,~T.; Krummel,~A.; Roberts,~S.
  Celebrating 25 years of {2D} {IR} spectroscopy. \emph{J. Chem. Phys.}
  \textbf{2024}, \emph{160}, 010401\relax
\mciteBstWouldAddEndPuncttrue
\mciteSetBstMidEndSepPunct{\mcitedefaultmidpunct}
{\mcitedefaultendpunct}{\mcitedefaultseppunct}\relax
\EndOfBibitem
\bibitem[Brixner \latin{et~al.}(2005)Brixner, Stenger, Vaswani, Cho,
  Blankenship, and Fleming]{brixner_two-dimensional_2005}
Brixner,~T.; Stenger,~J.; Vaswani,~H.~M.; Cho,~M.; Blankenship,~R.~E.;
  Fleming,~G.~R. Two-dimensional spectroscopy of electronic couplings in
  photosynthesis. \emph{Nature} \textbf{2005}, \emph{434}, 625--628\relax
\mciteBstWouldAddEndPuncttrue
\mciteSetBstMidEndSepPunct{\mcitedefaultmidpunct}
{\mcitedefaultendpunct}{\mcitedefaultseppunct}\relax
\EndOfBibitem
\bibitem[Rouxel \latin{et~al.}(2026)Rouxel, L{\"u}ttig, Jones, and
  Zigmantas]{rouxel_charting_2026}
Rouxel,~R.; L{\"u}ttig,~J.; Jones,~M.~R.; Zigmantas,~D. Charting light
  harvesting in purple bacteria in vivo. \emph{Proceedings of the National
  Academy of Sciences} \textbf{2026}, \emph{123}, e2537487123\relax
\mciteBstWouldAddEndPuncttrue
\mciteSetBstMidEndSepPunct{\mcitedefaultmidpunct}
{\mcitedefaultendpunct}{\mcitedefaultseppunct}\relax
\EndOfBibitem
\bibitem[Agathangelou \latin{et~al.}(2021)Agathangelou, Javed, Sessa, Solinas,
  Joffre, and Ogilvie]{agathangelou_phase-modulated_2021}
Agathangelou,~D.; Javed,~A.; Sessa,~F.; Solinas,~X.; Joffre,~M.; Ogilvie,~J.~P.
  Phase-modulated rapid-scanning fluorescence-detected two-dimensional
  electronic spectroscopy. \emph{The Journal of Chemical Physics}
  \textbf{2021}, \emph{155}, 094201\relax
\mciteBstWouldAddEndPuncttrue
\mciteSetBstMidEndSepPunct{\mcitedefaultmidpunct}
{\mcitedefaultendpunct}{\mcitedefaultseppunct}\relax
\EndOfBibitem
\bibitem[Goetz \latin{et~al.}(2018)Goetz, Li, Kolb, Pflaum, and
  Brixner]{goetz_coherent_2018}
Goetz,~S.; Li,~D.; Kolb,~V.; Pflaum,~J.; Brixner,~T. Coherent two-dimensional
  fluorescence micro-spectroscopy. \emph{Opt. Express} \textbf{2018},
  \emph{26}, 3915--3925, Number: 4\relax
\mciteBstWouldAddEndPuncttrue
\mciteSetBstMidEndSepPunct{\mcitedefaultmidpunct}
{\mcitedefaultendpunct}{\mcitedefaultseppunct}\relax
\EndOfBibitem
\bibitem[Tekavec \latin{et~al.}(2007)Tekavec, Lott, and
  Marcus]{tekavec_fluorescence-detected_2007}
Tekavec,~P.~F.; Lott,~G.~A.; Marcus,~A.~H. Fluorescence-detected
  two-dimensional electronic coherence spectroscopy by acousto-optic phase
  modulation. \emph{The Journal of Chemical Physics} \textbf{2007}, \emph{127},
  214307\relax
\mciteBstWouldAddEndPuncttrue
\mciteSetBstMidEndSepPunct{\mcitedefaultmidpunct}
{\mcitedefaultendpunct}{\mcitedefaultseppunct}\relax
\EndOfBibitem
\bibitem[Bakulin \latin{et~al.}(2016)Bakulin, Silva, and
  Vella]{bakulin_ultrafast_2016}
Bakulin,~A.~A.; Silva,~C.; Vella,~E. Ultrafast {Spectroscopy} with
  {Photocurrent} {Detection}: {Watching} {Excitonic} {Optoelectronic} {Systems}
  at {Work}. \emph{J. Phys. Chem. Lett.} \textbf{2016}, \emph{7},
  250--258\relax
\mciteBstWouldAddEndPuncttrue
\mciteSetBstMidEndSepPunct{\mcitedefaultmidpunct}
{\mcitedefaultendpunct}{\mcitedefaultseppunct}\relax
\EndOfBibitem
\bibitem[Uhl \latin{et~al.}(2021)Uhl, Bangert, Bruder, and
  Stienkemeier]{uhl_coherent_2021}
Uhl,~D.; Bangert,~U.; Bruder,~L.; Stienkemeier,~F. Coherent optical {2D}
  photoelectron spectroscopy. \emph{Optica, OPTICA} \textbf{2021}, \emph{8},
  1316--1324\relax
\mciteBstWouldAddEndPuncttrue
\mciteSetBstMidEndSepPunct{\mcitedefaultmidpunct}
{\mcitedefaultendpunct}{\mcitedefaultseppunct}\relax
\EndOfBibitem
\bibitem[Karki \latin{et~al.}(2014)Karki, Widom, Seibt, Moody, Lonergan,
  Pullerits, and Marcus]{karki_coherent_2014}
Karki,~K.~J.; Widom,~J.~R.; Seibt,~J.; Moody,~I.; Lonergan,~M.~C.;
  Pullerits,~T.; Marcus,~A.~H. Coherent two-dimensional photocurrent
  spectroscopy in a {PbS} quantum dot photocell. \emph{Nat. Commun.}
  \textbf{2014}, \emph{5}, 5869\relax
\mciteBstWouldAddEndPuncttrue
\mciteSetBstMidEndSepPunct{\mcitedefaultmidpunct}
{\mcitedefaultendpunct}{\mcitedefaultseppunct}\relax
\EndOfBibitem
\bibitem[L{\'o}pez-Ortiz \latin{et~al.}(2024)L{\'o}pez-Ortiz, Bolzonello,
  Bruschi, Fresch, Collini, Hu, Croce, van Hulst, and
  Gorostiza]{lopez-ortiz_photoelectrochemical_2024}
L{\'o}pez-Ortiz,~M.; Bolzonello,~L.; Bruschi,~M.; Fresch,~E.; Collini,~E.;
  Hu,~C.; Croce,~R.; van Hulst,~N.~F.; Gorostiza,~P. Photoelectrochemical
  {Two}-{Dimensional} {Electronic} {Spectroscopy} ({PEC2DES}) of {Photosystem}
  {I}: {Charge} {Separation} {Dynamics} {Hidden} in a {Multichromophoric}
  {Landscape}. \emph{ACS Appl. Mater. Interfaces} \textbf{2024}, \relax
\mciteBstWouldAddEndPunctfalse
\mciteSetBstMidEndSepPunct{\mcitedefaultmidpunct}
{}{\mcitedefaultseppunct}\relax
\EndOfBibitem
\bibitem[Roeding and Brixner(2018)Roeding, and Brixner]{roeding_coherent_2018}
Roeding,~S.; Brixner,~T. Coherent two-dimensional electronic mass spectrometry.
  \emph{Nature Communications} \textbf{2018}, \emph{9}, 2519\relax
\mciteBstWouldAddEndPuncttrue
\mciteSetBstMidEndSepPunct{\mcitedefaultmidpunct}
{\mcitedefaultendpunct}{\mcitedefaultseppunct}\relax
\EndOfBibitem
\bibitem[Bruder \latin{et~al.}(2018)Bruder, Bangert, Binz, Uhl, Vexiau,
  Bouloufa-Maafa, Dulieu, and Stienkemeier]{bruder_coherent_2018}
Bruder,~L.; Bangert,~U.; Binz,~M.; Uhl,~D.; Vexiau,~R.; Bouloufa-Maafa,~N.;
  Dulieu,~O.; Stienkemeier,~F. Coherent multidimensional spectroscopy of dilute
  gas-phase nanosystems. \emph{Nat. Commun.} \textbf{2018}, \emph{9}, 4823,
  Number: 1\relax
\mciteBstWouldAddEndPuncttrue
\mciteSetBstMidEndSepPunct{\mcitedefaultmidpunct}
{\mcitedefaultendpunct}{\mcitedefaultseppunct}\relax
\EndOfBibitem
\bibitem[Solowan \latin{et~al.}(2022)Solowan, Mal{\'y}, and
  Brixner]{solowan_direct_2022}
Solowan,~H.-P.; Mal{\'y},~P.; Brixner,~T. Direct comparison of molecular-beam
  vs liquid-phase pump{\textendash}probe and two-dimensional spectroscopy on
  the example of azulene. \emph{J. Chem. Phys.} \textbf{2022}, \emph{157},
  044201\relax
\mciteBstWouldAddEndPuncttrue
\mciteSetBstMidEndSepPunct{\mcitedefaultmidpunct}
{\mcitedefaultendpunct}{\mcitedefaultseppunct}\relax
\EndOfBibitem
\bibitem[Ma \latin{et~al.}(2023)Ma, Chen, Xu, and
  Fournier]{ma_two-dimensional_2023}
Ma,~Z.; Chen,~L.; Xu,~C.; Fournier,~J.~A. Two-{Dimensional} {Infrared}
  {Spectroscopy} of {Isolated} {Molecular} {Ions}. \emph{J. Phys. Chem. Lett.}
  \textbf{2023}, \emph{14}, 9683--9689\relax
\mciteBstWouldAddEndPuncttrue
\mciteSetBstMidEndSepPunct{\mcitedefaultmidpunct}
{\mcitedefaultendpunct}{\mcitedefaultseppunct}\relax
\EndOfBibitem
\bibitem[Dutta \latin{et~al.}(2025)Dutta, Ma, Fournier, and
  Reppert]{dutta_nonlinear_2025}
Dutta,~R.; Ma,~Z.; Fournier,~J.~A.; Reppert,~M. Nonlinear response from linear
  oscillators: {Gas} phase {2D} action spectroscopy. \emph{J. Chem. Phys.}
  \textbf{2025}, \emph{163}, 054113\relax
\mciteBstWouldAddEndPuncttrue
\mciteSetBstMidEndSepPunct{\mcitedefaultmidpunct}
{\mcitedefaultendpunct}{\mcitedefaultseppunct}\relax
\EndOfBibitem
\bibitem[Xie and Xu(2022)Xie, and Xu]{xie_fourier-transform_2022}
Xie,~Q.; Xu,~X.~G. Fourier-{Transform} {Atomic} {Force} {Microscope}-{Based}
  {Photothermal} {Infrared} {Spectroscopy} with {Broadband} {Source}.
  \emph{Nano Lett.} \textbf{2022}, \emph{22}, 9174--9180\relax
\mciteBstWouldAddEndPuncttrue
\mciteSetBstMidEndSepPunct{\mcitedefaultmidpunct}
{\mcitedefaultendpunct}{\mcitedefaultseppunct}\relax
\EndOfBibitem
\bibitem[Xie \latin{et~al.}(2024)Xie, Zhang, Janzen, Edgar, and
  Xu]{xie_atomic-force-microscopy-based_2024}
Xie,~Q.; Zhang,~Y.; Janzen,~E.; Edgar,~J.~H.; Xu,~X.~G.
  Atomic-force-microscopy-based time-domain two-dimensional infrared
  nanospectroscopy. \emph{Nat. Nanotechnol.} \textbf{2024}, \emph{19},
  1108--1115\relax
\mciteBstWouldAddEndPuncttrue
\mciteSetBstMidEndSepPunct{\mcitedefaultmidpunct}
{\mcitedefaultendpunct}{\mcitedefaultseppunct}\relax
\EndOfBibitem
\bibitem[Tiwari \latin{et~al.}(2018)Tiwari, Matutes, Gardiner, Jansen, Cogdell,
  and Ogilvie]{tiwari_spatially-resolved_2018}
Tiwari,~V.; Matutes,~Y.~A.; Gardiner,~A.~T.; Jansen,~T. L.~C.; Cogdell,~R.~J.;
  Ogilvie,~J.~P. Spatially-resolved fluorescence-detected two-dimensional
  electronic spectroscopy probes varying excitonic structure in photosynthetic
  bacteria. \emph{Nat. Commun.} \textbf{2018}, \emph{9}, 4219\relax
\mciteBstWouldAddEndPuncttrue
\mciteSetBstMidEndSepPunct{\mcitedefaultmidpunct}
{\mcitedefaultendpunct}{\mcitedefaultseppunct}\relax
\EndOfBibitem
\bibitem[Jana \latin{et~al.}(2024)Jana, Durst, and
  Lippitz]{jana_fluorescence-detected_2024}
Jana,~S.; Durst,~S.; Lippitz,~M. Fluorescence-detected two-dimensional
  electronic spectroscopy of a single molecule. \emph{Nano Lett.}
  \textbf{2024}, \emph{24}, 12576--12581\relax
\mciteBstWouldAddEndPuncttrue
\mciteSetBstMidEndSepPunct{\mcitedefaultmidpunct}
{\mcitedefaultendpunct}{\mcitedefaultseppunct}\relax
\EndOfBibitem
\bibitem[Gr{\'e}goire \latin{et~al.}(2017)Gr{\'e}goire, Srimath~Kandada, Vella,
  Tao, Leonelli, and Silva]{gregoire_incoherent_2017}
Gr{\'e}goire,~P.; Srimath~Kandada,~A.~R.; Vella,~E.; Tao,~C.; Leonelli,~R.;
  Silva,~C. Incoherent population mixing contributions to phase-modulation
  two-dimensional coherent excitation spectra. \emph{The Journal of Chemical
  Physics} \textbf{2017}, \emph{147}, 114201\relax
\mciteBstWouldAddEndPuncttrue
\mciteSetBstMidEndSepPunct{\mcitedefaultmidpunct}
{\mcitedefaultendpunct}{\mcitedefaultseppunct}\relax
\EndOfBibitem
\bibitem[Bolzonello \latin{et~al.}(2023)Bolzonello, Bruschi, Fresch, and van
  Hulst]{bolzonello_nonlinear_2023}
Bolzonello,~L.; Bruschi,~M.; Fresch,~B.; van Hulst,~N.~F. Nonlinear {Optical}
  {Spectroscopy} of {Molecular} {Assemblies}: {What} {Is} {Gained} and {Lost}
  in {Action} {Detection}? \emph{J. Phys. Chem. Lett.} \textbf{2023},
  \emph{14}, 11438--11446\relax
\mciteBstWouldAddEndPuncttrue
\mciteSetBstMidEndSepPunct{\mcitedefaultmidpunct}
{\mcitedefaultendpunct}{\mcitedefaultseppunct}\relax
\EndOfBibitem
\bibitem[{Leonas Valkunas, Darius Abramavicius, Tom{\'a}{\v s} Man{\v
  c}al}(2013)]{leonas_valkunas_darius_abramavicius_tomas_mancal_molecular_2013}
{Leonas Valkunas, Darius Abramavicius, Tom{\'a}{\v s} Man{\v c}al},
  \emph{Molecular {Excitation} {Dynamics} and {Relaxation}}, 1st ed.;
  WILEY-VCH: Weinheim, 2013\relax
\mciteBstWouldAddEndPuncttrue
\mciteSetBstMidEndSepPunct{\mcitedefaultmidpunct}
{\mcitedefaultendpunct}{\mcitedefaultseppunct}\relax
\EndOfBibitem
\bibitem[Man{\v c}al and collective(2026)Man{\v c}al, and
  collective]{mancal_quantarhei_2026}
Man{\v c}al,~T.; collective,~a. {QuantaRhei}. 2026;
  \url{https://github.com/tmancal74/quantarhei}\relax
\mciteBstWouldAddEndPuncttrue
\mciteSetBstMidEndSepPunct{\mcitedefaultmidpunct}
{\mcitedefaultendpunct}{\mcitedefaultseppunct}\relax
\EndOfBibitem
\bibitem[Tan(2008)]{tan_theory_2008}
Tan,~H.-S. Theory and phase-cycling scheme selection principles of collinear
  phase coherent multi-dimensional optical spectroscopy. \emph{The Journal of
  Chemical Physics} \textbf{2008}, \emph{129}, 124501, Number: 12\relax
\mciteBstWouldAddEndPuncttrue
\mciteSetBstMidEndSepPunct{\mcitedefaultmidpunct}
{\mcitedefaultendpunct}{\mcitedefaultseppunct}\relax
\EndOfBibitem
\bibitem[Mal{\'y} \latin{et~al.}(2020)Mal{\'y}, L{\"u}ttig, Mueller, Schreck,
  Lambert, and Brixner]{maly_coherently_2020}
Mal{\'y},~P.; L{\"u}ttig,~J.; Mueller,~S.; Schreck,~M.~H.; Lambert,~C.;
  Brixner,~T. Coherently and fluorescence-detected two-dimensional electronic
  spectroscopy: direct comparison on squaraine dimers. \emph{Phys. Chem. Chem.
  Phys.} \textbf{2020}, \emph{22}, 21222--21237\relax
\mciteBstWouldAddEndPuncttrue
\mciteSetBstMidEndSepPunct{\mcitedefaultmidpunct}
{\mcitedefaultendpunct}{\mcitedefaultseppunct}\relax
\EndOfBibitem
\bibitem[Rueda~Espinosa \latin{et~al.}(2026)Rueda~Espinosa, Xu, and
  Kananenka]{rueda_espinosa_theory_2026}
Rueda~Espinosa,~K.~J.; Xu,~X.~G.; Kananenka,~A.~A. Theory and {Simulations} of
  {Atomic}-force-microscopy based {Two}-dimensional {Electronic}
  {Spectroscopy}. 2026;
  \url{https://chemrxiv.org/doi/full/10.26434/chemrxiv.15005099/v1}\relax
\mciteBstWouldAddEndPuncttrue
\mciteSetBstMidEndSepPunct{\mcitedefaultmidpunct}
{\mcitedefaultendpunct}{\mcitedefaultseppunct}\relax
\EndOfBibitem
\bibitem[Mal{\'y} \latin{et~al.}(2023)Mal{\'y}, L{\"u}ttig, Rose, Turkin,
  Lambert, Krich, and Brixner]{maly_separating_2023}
Mal{\'y},~P.; L{\"u}ttig,~J.; Rose,~P.~A.; Turkin,~A.; Lambert,~C.;
  Krich,~J.~J.; Brixner,~T. Separating single- from multi-particle dynamics in
  nonlinear spectroscopy. \emph{Nature} \textbf{2023}, \emph{616},
  280--287\relax
\mciteBstWouldAddEndPuncttrue
\mciteSetBstMidEndSepPunct{\mcitedefaultmidpunct}
{\mcitedefaultendpunct}{\mcitedefaultseppunct}\relax
\EndOfBibitem
\bibitem[L{\"u}ttig and Mal{\'y}(2026)L{\"u}ttig, and
  Mal{\'y}]{luttig_simple_2026}
L{\"u}ttig,~J.; Mal{\'y},~P. A simple method to extract nonlinear signals in
  time-resolved spectroscopy. \emph{J. Phys. B: At. Mol. Opt. Phys.}
  \textbf{2026}, \emph{59}, 063001\relax
\mciteBstWouldAddEndPuncttrue
\mciteSetBstMidEndSepPunct{\mcitedefaultmidpunct}
{\mcitedefaultendpunct}{\mcitedefaultseppunct}\relax
\EndOfBibitem
\bibitem[Mal{\'y} and Man{\v c}al(2018)Mal{\'y}, and Man{\v
  c}al]{maly_signatures_2018}
Mal{\'y},~P.; Man{\v c}al,~T. Signatures of exciton delocalization and
  exciton{\textendash}exciton annihilation in fluorescence-detected
  two-dimensional coherent spectroscopy. \emph{J. Phys. Chem. Lett.}
  \textbf{2018}, \emph{9}, 5654--5659\relax
\mciteBstWouldAddEndPuncttrue
\mciteSetBstMidEndSepPunct{\mcitedefaultmidpunct}
{\mcitedefaultendpunct}{\mcitedefaultseppunct}\relax
\EndOfBibitem
\bibitem[Kunsel \latin{et~al.}(2019)Kunsel, Tiwari, Matutes, Gardiner, Cogdell,
  Ogilvie, and Jansen]{kunsel_simulating_2019}
Kunsel,~T.; Tiwari,~V.; Matutes,~Y.~A.; Gardiner,~A.~T.; Cogdell,~R.~J.;
  Ogilvie,~J.~P.; Jansen,~T. L.~C. Simulating {Fluorescence}-{Detected}
  {Two}-{Dimensional} {Electronic} {Spectroscopy} of {Multichromophoric}
  {Systems}. \emph{The Journal of Physical Chemistry B} \textbf{2019},
  \emph{123}, 394--406\relax
\mciteBstWouldAddEndPuncttrue
\mciteSetBstMidEndSepPunct{\mcitedefaultmidpunct}
{\mcitedefaultendpunct}{\mcitedefaultseppunct}\relax
\EndOfBibitem
\end{mcitethebibliography}

\end{document}